**Entre el pacifismo y la energía nuclear (1930-1955)**[*]

Diego H. de Mendoza
Centro de Estudios de Historia de la Ciencia José Babini
Escuela de Humanidades - UNSAM

A mediados de 1929 la caída de la bolsa de New York inició un largo período de depresión económica mundial. El miedo al comunismo impulsó a muchos de los más importantes industriales alemanes a apoyar el ascenso político de los nazis. Durante estos años Albert Einstein, que vivía en Berlín desde 1914, apareció como portavoz de un pacifismo radical que promovió la resistencia individual a la guerra y el principio de desarme universal. Estas ideas fueron expresadas por Einstein en numerosas cartas, entrevistas y manifiestos que circularon entonces por Europa con su firma. En 1930, uno de estos manifiestos sostenía:

> ¡Naciones enteras están en peligro! / ¿Usted conoce el significado de una nueva guerra, la cual podría usar los medios de destrucción que la ciencia está incesantemente perfeccionando? / ¿Usted sabe que en el futuro la guerra no será ya más beneficiosa para nadie, ya que no sólo depósitos de armas, municiones o alimentos, sino también todos los centros industriales importantes serán el blanco de los ataques? (citado en Pais, 1994: 175).

Durante los inviernos de 1930-31 y el siguiente Einstein visitó el California Institute of Technology, en Pasadena, invitado por su director Robert Millikan. En una de sus conferencias pronunciadas allí, en febrero de 1931, Einstein se preguntaba: "¿Por qué esta magnificente ciencia aplicada que ahorra trabajo y hace la vida más fácil nos trajo tan poca felicidad?" Y Agregaba:

> La respuesta más simple es: porque no hemos aprendido a hacer un uso sensible de ella. / En la guerra esta sirve para que nos podamos envenenar y mutilar unos a otros. En la paz esta ha hecho nuestras vidas más ajetreadas e inciertas. En lugar de liberarnos en gran medida del trabajo espiritualmente agotador, ha hecho de los hombres esclavos de las máquinas, quienes en su mayor parte completan sus monótonas y extensas jornadas de trabajo con disgusto y deben temblar continuamente por sus pobres raciones (*New York Times*, 1931).

También a mediados de 1931, en la Conferencia Internacional de Opositores a la Guerra, hizo un llamado a los científicos a rechazar toda cooperación en la investigación de nuevos instrumentos de guerra. En noviembre, anticipando la Conferencia sobre Desarme en febrero del siguiente año, Einstein sostuvo que si bien existen la Liga de las Naciones y la Corte Mundial, la primera no es mucho más que "un lugar de encuentro" y la segunda "no tiene los medios de hacer cumplir sus decisiones". Así, estas instituciones "son incapaces de dar seguridad a ningún estado en caso de que fuera atacado". Y concluyó:

> Si no acordamos en limitar la soberanía de los Estados individuales, y si, al mismo tiempo, todos los Estados no garantizan realizar una acción común contra cualquiera de ellos que abierta o subrepticiamente viole una decisión de la Corte Mundial, no podremos escapar de la situación de general anarquía y amenaza.

Dado que los meros acuerdos de reducción de armamento no garantizan nada, argumentó Einstein, es necesario que una corte de arbitraje obligatorio tenga a su disposición la autoridad ejecutiva y el respaldo de todos los Estados participantes para tomar sanciones militares y económicas contra los perturbadores de la paz. Finalmente, en esta misma intervención atacó el nacionalismo exacerbado, "que recibe el atractivo pero erróneo nombre de patriotismo", como el

---

"mayor obstáculo para el orden internacional" y como instrumento para justificar el servicio militar obligatorio: "Los objetores de conciencia deben ser especialmente protegidos por un acuerdo internacional" (Einstein, 1931). En cuanto al rol de los gobiernos y las naciones, a comienzos de 1932, sugirió que Estados Unidos, Inglaterra, Alemania y Francia promovieran un boicot económico a Japón como intimación para que este país ponga fin a su intervención militar en China:

> ¿Ustedes creen que Japón encontrará un gobierno que desee tomar la responsabilidad de hundir a su país en tan peligrosa aventura [...] ¿Por qué no se hace esto? ¿Por qué cada persona y cada nación deben temblar por su existencia? Porque cada uno busca su miserable, momentánea ventaja y no se subordina a sí mismo por el bien y la prosperidad de la comunidad (*New York Times*, 1932).

A fines de marzo de 1932, Einstein viajó a Oxford, Inglaterra, donde recibió la invitación para formar parte del recién creado Instituto de Estudios Avanzados en New Jersey, Estados Unidos. Si bien había rechazado una oferta de trasladarse a Estados Unidos en 1927, esta vez, la tensa situación política de Alemania lo persuadió de considerar esta posibilidad.

**Ascenso del nazismo y cambio de postura**

Ya a comienzos de 1920 había sido montada una campaña de reacción contra la teoría general de la relatividad por un grupo de físicos alemanes, entre los que se encontraban los premios Nobel de física Philipp Lenard y Johannes Stark, y Ernst Gehrcke, experto en óptica experimental y profesor del Physikalisch-Technische Reichsanstalt. Este grupo de científicos no solamente se ubicaba políticamente en la extrema derecha, sino que también adhería a valores conservadores en cuanto a lo que consideraban conocimiento científico legítimo, como el mecanicismo clásico, la causalidad, la visualización y la preferencia por el experimento a la teoría. Este grupo rechazó las teorías de Einstein por considerarlas absurdas, espiritualmente peligrosas y sin fundamento experimental concluyente. También sostuvieron que los resultados de las ecuaciones de Einstein que habrían sido supuestamente confirmados podían obtenerse a partir de la física clásica (Kragh, 1999: 103-104; Wazeck 2004). Lo cierto es que durante el período conocido como la república de Weimar la actividad de los "antirelativistas" no evitó la creciente difusión de la relatividad ni de las nuevas ideas de la física cuántica, a las que también se oponían, y derivó en la propia marginación de este grupo, que canalizó la falta de reconocimiento hacia sentimientos antidemocráticos y antisemitas. En 1924, mientras Hitler estuvo en prisión por su intervención en el fallido golpe de estado en Munich, Lenard y Stark manifestaron su apoyo al líder del Partido Nacionalsocialista. Este respaldo de dos científicos de renombre internacional en momentos de zozobra política fue apreciado por Hitler, quien, una vez en el poder, iba a ayudar a que las ideas de este grupo retornaran con renovado vigor y tomaran la forma de lo que se llamó "física aria" (Walker, 1989: 64-65).

Cuando en enero de 1933 Hitler se transformó en canciller de Alemania y en marzo obtuvo poderes dictatoriales, Einstein se encontraba en Pasadena. En una dura declaración pública anunció desde allí que no regresaría a Alemania y a fines de marzo renunció a la Academia Prusiana de Ciencias. Los nazis confiscaron su cuenta bancaria, la caja de seguridad de su esposa y su casa de descanso en Caputh, cerca de Berlín. En este clima enrarecido, es importante señalar que ni Max Planck, ni Max von Laue, ni Walther Nernst fueron intimidados por los opositores de Einstein. El 2 de mayo de 1933, por ejemplo, Planck hizo una declaración en la que afirmó que "Einstein no es sólo uno de nuestros muchos físicos de talla; es, además, el físico con cuyas obras, publicadas por nuestra academia, la física ha experimentado un progreso cuya importancia sólo puede compararse con los avances logrados por Johannes Kepler e Isaac Newton [...]" (citado en Hoffmann, 1985: 150).

Durante esos días, Einstein viajó a Bélgica, al pueblo de Le Coq-sur-Mer. Allí, por orden del rey Alberto, permaneció protegido por guardaespaldas. Una declaración del 20 de julio muestra un giro radical en la postura pacifista de Einstein. Cuando se le pidió que hablara en favor de dos objetores de conciencia belgas, sostuvo Einstein:

> Imaginen a Bélgica ocupada por la Alemania actual. Las cosas serían mucho peor que en 1914, y entonces no fueron nada buenas. Por eso tengo que decir con toda franqueza: si yo fuera belga, y dadas las actuales circunstancias, no me negaría a prestar el servicio militar; por el contrario, entraría en dicha organización con alegría y pensando que de esa manera contribuiría a salvar a la civilización europea (citado en Hoffmann, 1985: 153).

Su súbito cambio de opinión, que lo llevaba ahora a sostener que no había que resistir más a la guerra y que había que promover el rearme de las naciones occidentales, causó consternación y motivó duras críticas en los círculos pacifistas intransigentes. En enero de 1935, en el diario *New York Sun*, Einstein argumentó sobre sus nuevas ideas. Hasta unos pocos años atrás, el rechazo de las armas parecía ir en favor del ideal de una organización supranacional, "sin embargo, este curso de acción no puede ser ya más recomendado, por lo menos para los países europeos". Si bien el rechazo del servicio militar por parte de un gran número de ciudadanos pudo considerarse como "una política constructiva" que apuntó a un genuino pacifismo y puso en evidencia "los aspectos inmorales del servicio militar compulsivo", aclaró Einstein: "Hoy, sin embargo, debemos reconocer que varias naciones poderosas hacen imposible para sus ciudadanos adoptar una posición política independiente". En estos países, funciona "una prensa esclavizada, un servicio de radio centralizado y un sistema educativo orientado hacia una política externa agresiva" y "el rechazo del servicio militar significa el martirio y la muerte". Como contrapartida, "en aquellos países que respetan los derechos políticos de sus ciudadanos, el rechazo del servicio militar sirve probablemente para debilitar la habilidad de los sectores saludables del mundo civilizado de resistir la agresión". La conclusión es que "hoy, ninguna persona inteligente debería apoyar la política de rechazar el servicio militar, al menos no en Europa, la cual está especialmente amenazada" (citado en Nathan y Norden, 1968: 254).

**La fisión del uranio**

En 1932 una nueva partícula elemental, que fue llamada neutrón, fue identificada por el británico James Chadwick del Laboratorio Cavendish, Cambridge. En seguida, los físicos y químicos comprendieron que esta nueva partícula, dado que carecía de carga eléctrica, podría ser utilizada como un eficiente proyectil contra el núcleo del átomo y, por lo tanto, para el estudio de las reacciones nucleares. En la Universidad de Roma de la Italia fascista, el físico Enrico Fermi y su grupo comenzaron a bombardear núcleos atómicos con neutrones. Al bombardear uranio, Fermi pensó que había logrado producir los primeros elementos con mayor peso atómico que el uranio, los llamados transuránidos. El anuncio de sus resultados, en 1934, empujó a Otto Hahn y Lisa Meitner, ambos del Instituto de Química Kaiser Wilhelm, a iniciar investigaciones en esta línea. Sin embargo, cuando Alemania invadió Austria, Lisa Meitner, de descendencia judía que había confiado en que su nacionalidad austríaca la mantendría protegida, debió escapar de la Gestapo. Con la ayuda de Niels Bohr, Meitner continuó sus investigaciones en Estocolmo. En septiembre de 1938, Mussolini introdujo las leyes antisemitas en Italia y Fermi, cuya esposa era judía, debió marcharse a los Estados Unidos, donde fue contratado como profesor en la Universidad de Columbia.

En el otoño de 1938, Otto Hahn y Fritz Strassman, en Berlín, comenzaron un cuidadoso análisis químico de los elementos producidos en colisiones de neutrones con núcleos de uranio. Para asombro de los autores, entre los productos resultantes identificaron el bario, un elemento cuyos átomos pesan aproximadamente la mitad que los de uranio. Hasta ese momento, las reacciones nucleares nunca habían producido elementos de pesos muy diferentes a los pesos originales de los elementos bombardeados. Con cierta vacilación, el 6 de enero de 1939, sin arriesgar una explicación, Hahn y Strassman publicaron estos resultados en la revista alemana *Die Naturwissenschaften* (vol. 27), aclarando que sus resultados estaban "en oposición a todos los fenómenos observados hasta el presente en física nuclear".

El análisis teórico de este fenómeno fue hecho por Lisa Meitner y su sobrino Otto Frisch, que integraba el equipo de Bohr en Copenhague. Si el núcleo atómico fuera un bloque monolítico, como sostenía Ernest Rutherford, razonaron, entonces no se podría dividir. Pero si se comporta como una gota de agua—modelo que había sido propuesto por Bohr—, entonces podría escindirse en dos partes aproximadamente iguales. Frisch preguntó a un biólogo qué término utilizaba para la división de las células. La respuesta del biólogo, "fisión", fue introducida en el artículo publicado en *Nature* (vol. 143), donde Meitner y Frisch explicaron la fragmentación violenta del núcleo de uranio. Dado que la suma de los pesos de los productos de la fisión eran menores que el peso del núcleo original, la pérdida de masa, especularon los autores, debe haberse transformado en la energía liberada durante el proceso de fisión. La célebre ecuación $E = mc^2$, publicada por Einstein en 1905, encontraba en este proceso su primera aplicación.

En este punto, los hechos se precipitan. Bohr se enteró de la novedad por Frisch y dos días más tarde viajó a los Estados Unidos, donde anunció la fisión nuclear a los físicos de aquel país. En abril de 1939 el equipo del Collège de France, liderado por Frédéric Joliot y su esposa Irène Curie—hija de Pierre y Marie Curie—, encontró que en cada fisión se emiten 3,5 neutrones, lo que demostraba que la reacción en cadena es posible. No hizo falta mucha especulación posterior para descubrir que este fenómeno podría utilizarse para construir explosivos con capacidad destructiva inédita. Una semana más tarde, refiriéndose a una reunión de la American Physical Society en el Bureau of Standards, Washington, de la cual participaron Niels Bohr y John Wheeler y a la posibilidad de que los experimentos que se estaban realizando sobre la fisión del núcleo de uranio pudieran producir explosiones no controladas, el periódico *Washington Post* publicó un artículo en donde se discute si uno de los experimentos que se están realizando podría "producir un cráter de dos millas" (*Washington Post*, 1939). Finalmente, a fines de junio el círculo se cerraba y la fisión del uranio 235 resultaba develada cuando Bohr y Wheeler enviaron a la revista norteamericana *Physical Review* el artículo titulado "Mecanismo de la fisión nuclear", que fue publicado el 1° de septiembre (vol. 56), sobre el inicio de la guerra.

Mientras esto ocurría, los intereses científicos de Einstein iban en otra dirección. En un artículo titulado "Einstein ve cercana la clave del universo", publicado en *New York Times* en ocasión de su sexagésimo cumpleaños, en marzo de 1939, Einstein afirmaba que "al presente la física teórica todavía está lejos de proveer un fundamento unificado sobre el cual podría basarse el tratamiento teórico de todos los fenómenos". Y agregaba:

> Tenemos una teoría general de la relatividad de los fenómenos macroscópicos, la cual, sin embargo, hasta la fecha es incapaz de dar cuenta de la estructura atómica de la materia y de los efectos cuánticos; y tenemos una teoría cuántica que es capaz de dar cuenta satisfactoriamente de un gran número de fenómenos atómicos y cuánticos, pero que, por su propia naturaleza, no se ajusta a los principios de relatividad.

En cuanto al estado de sus investigaciones en el terreno de la teoría del campo unificado, que había iniciado veinte años atrás, comentaba que desde la formulación de la teoría general de la relatividad ha existido el problema de expresar "bajo un mismo concepto matemático unificado el campo gravitatorio, el campo electromagnético y las partículas materiales". Y sostenía que el año anterior había descubierto una nueva solución que ahora él, junto con dos colaboradores, estaba intentando desarrollar hasta el punto en que pudiera ser puesta a prueba experimentalmente. Cuando en la misma entrevista se le preguntó sobre su opinión acerca de la liberación de energía por el átomo de uranio, Einstein respondió que "los resultados referidos a la partición del átomo hasta ahora no justifican la creencia de un uso práctico de la energía atómica liberada en este proceso" (Laurence, 1939). Esto ocurría en marzo. Pocos meses después, en agosto de 1939, Einstein ya había cambiado de opinión.

**La caja de Pandora**

Existe una notable cantidad de bibliografía acerca del programa norteamericano que logró construir las primeras bombas atómicas, conocido como Proyecto Manhattan. Muchos historiadores suelen asumir que el comienzo de este programa—que por la magnitud de recursos materiales y por el número de científicos, ingenieros y técnicos involucrados marcó un hito en la historia del desarrollo científico y tecnológico—se ubica en la carta que Einstein dirigió al presidente norteamericano

Franklin D. Roosevelt a mediados de 1939. La historia de esta carta está vinculada al llamado grupo de "los húngaros"—en referencia a los físicos Leo Szilard, Eugene Wigner y Edward Teller, húngaros de origen con ciudadanía alemana—. Szilard visitó en dos ocasiones Peconic, Long Island, donde Einstein tomaba sus vacaciones. Durante la primera visita, el 15 de julio, fue acompañado por Wigner. Estos dos físicos pusieron al tanto a Einstein sobre las investigaciones en fisión. "Einstein estaba inmerso en su trabajo y no seguía los últimos desarrollos de la física", dirá más tarde Wigner. También le contaron sobre el papel estratégico del uranio, la alta probabilidad de que los alemanes pudieran producir bombas atómicas y sobre la necesidad de que el gobierno norteamericano tomara medidas urgentes. Szilard recuerda que Einstein comprendió inmediatamente las consecuencias de una reacción en cadena. En esa misma visita, Einstein entregó a Wigner el borrador de una carta en alemán con la tarea de mejorar el texto. El plan original era convencer a un ministro belga sobre la necesidad de privar de uranio a los nazis. A los pocos días de esta visita Szilard cambió de opinión y llamó por teléfono a Einstein para consultarle su cambio de planes: incitar al presidente de los Estados Unidos a construir la bomba. Szilard hizo una segunda visita a Peconic, ahora acompañado por Teller, el mismo que una década más tarde se transformaría en "el padre de la bomba H". Szilard y Teller llevaron a Eintsein dos posibles versiones manuscritas dirigidas a Roosevelt para que Einstein decida. Finalmente, decidieron que se le enviarían los dos textos mecanografiados para que Einstein pudiera decidir con calma cuál firmar (Closets, 2005: 24-33; 49-61; Walsh, 1973: 530).

En la versión final de la carta seleccionada, Einstein sostenía que tenía razones para creer que "el elemento uranio puede transformarse en una nueva e importante fuente de energía en el futuro inmediato [...] Parece ahora cierto que esta [reacción en cadena] podría ser realizada en el futuro inmediato". Además, este nuevo fenómeno "podría conducir a la construcción de bombas y es concebible—si bien mucho menos cierto—que bombas extremadamente poderosas de un nuevo tipo puedan ser construidas. Una simple bomba de este tipo, transportada por barco y hecha explotar en un puerto podría destruir completamente el puerto junto con el territorio circundante. Sin embargo, tales bombas podrían ser muy pesadas para ser transportadas por aire". Einstein afirmaba que "Alemania ha suspendido las ventas de uranio de las minas de Checoslovaquia, de las cuales se ha apoderado". Finalmente, menciona al físico Weizsäcker, quien trabaja en el Instituto Kaiser Wilhelm de Berlín, "donde se han reproducido varios trabajos norteamericanos sobre el uranio", razón por la cual alertaba a Roosevelt sobre la necesidad de tomar algún tipo de iniciativa. La carta se hizo llegar a Alexander Sachs, quien había ayudado a Szilard en la elaboración de las versiones enviadas a Einstein. Sach, economista y consejero privado de Roosevelt, se encargó de poner la carta en manos de éste el 3 de octubre (la carta es reproducida en Nathan y Norden, 1968: 295). El 19 de octubre Roosevelt respondió a Einstein con una carta de agradecimiento y creó el Advisory Committee on Uranium, el cual, sin embargo, no desplegó una gran actividad y no tuvo efectos determinantes. Al mes siguiente estalló la guerra. Entonces se le sugirió a Einstein que escribiera una segunda carta de la que se enterara el presidente. Con fecha 7 de marzo de 1940, en esta segunda carta Einstein sostenía:

> El último año, cuando comprendí que resultados de importancia nacional podrían surgir de la investigación con uranio, pensé que mi deber era informar a la Administración de esta posibilidad [...] Desde el estallido de la guerra, el interés en el uranio se ha intensificado en Alemania. Yo ahora me he enterado que allí se están llevando a cabo investigaciones en estricto secreto (citado en Nathan y Norden, 1968: 299).

En mayo de 1940, un grupo de diecisiete científicos de la Universidad de Princeton envió un telegrama al presidente Roosevelt declarando que los Estados Unidos deberían asistir a los aliados. Entre los diecisiete firmantes del telegrama se encontraban Einstein, Wigner y Herman Weyl. El mensaje clamaba un "enfático desacuerdo" con la petición enviada al presidente de los Estados Unidos por los 500 miembros de la American Association of Scientific Workers, en donde se pedía el mantenimiento de la neutralidad como forma de preservar la paz en los Estados Unidos. La mejor "defensa nacional consiste en asistir a aquellas fuerzas" que se oponen a "la agresión totalitaria" (*New York Times*, 1940; *Washington Post*, 1940).

Fueron estas intervenciones la causa de que posteriormente se responsabilizara a Einstein de haber dado el impulso inicial al Proyecto Manhattan. Años más tarde, Einstein repetiría en varias ocasiones que si hubiera sabido que los alemanes no tendrían éxito en producir su bomba atómica, "no habría movido un dedo". Sin embargo, entre la célebre carta de Einstein a Roosevelt y la puesta en marcha del Proyecto Manhattan tuvieron lugar varios acontecimientos que parecen diluir la responsabilidad directa de Einstein. Fue recién en 1941 que el interés de los Estados Unidos por el desarrollo de una bomba atómica comenzó a volverse realidad, junto con la decisión de iniciar los preparativos para su entrada en la guerra (Kragh, 1999: 265-66). Entre otras iniciativas, se creó entonces la Office of Scientific Reserch and Development, bajo la dirección del físico e ingeniero Vannevar Bush—quien al final de la guerra aparecía en su país como el "icono científico" más conocido después de Einstein (Blanpied, 1998: 35)—. Es en este momento que los físicos, químicos e ingenieros norteamericanos comenzaron a trabajar en la reacción en cadena del uranio y es luego del ataque japonés a Pearl Harbor, el 7 de diciembre, que el programa nuclear norteamericano se expandió a escalas sin precedentes. El historiador Stanley Goldberg argumenta de forma convincente en contra de la versión aceptada que sostiene que la decisión de construir una bomba atómica fue el producto del consenso alcanzado en base al análisis de estudios técnicos. Contra esta versión, él sostiene que "la decisión fue tomada en abril de 1941 por Vannevar Bush" (Goldberg, 1992: 429), a quien en noviembre de 1941 el editor científico del *New York Times* Waldemar Kaempffert caracterizaba como "el zar de la investigación" (Kevles, 1995: 300). Como parte del mismo escenario, digamos que Einstein fue excluido de toda participación en estas iniciativas, si bien hay evidencias que sugieren que Einstein fue contactado directamente por algunos científicos que integraban el proyecto para requerir su ayuda sobre problemas de índole teórica (Schwartz, 1989: 281).

En junio de 1943 Einstein firmó un contrato con la U.S. Navy Bureau of Ordnance para desempeñarse como consultor. En marzo de 1944, el *New York Times* citaba a *Star Shell*, una publicación de aquella institución, donde se mencionaba que Einstein estaba trabajando "en teoría de explosiones, buscando determinar qué leyes gobiernan las más oscuras ondas de la detonación, por qué ciertas explosiones tienen marcados efectos direccionales y otras teorías altamente técnicas", aunque esto en nada se vinculaba con la construcción de un artefacto atómico. También agrega la misma nota que la Marina no le ha pedido "que se cortara el pelo o que usara uniforme" (Schwarz, 1944: SM16).

En cuanto al proyecto de bomba atómica de la Alemania nazi, que había motivado la intervención de Einstein, puede decirse que en 1940 Werner Heisenberg—líder de las investigaciones atómicas desde Berlín—intentó construir un reactor nuclear. Ante la encrucijada sobre si el grafito o el agua pesada era el elemento adecuado para actuar de moderador, el profesor Bothe, de la Universidad de Heidelberg, se equivocó y aconsejó agua pesada. Esta decisión le costó caro al programa alemán. En los Estados Unidos, Leo Szilard llegó a la conclusión contraria. Por su parte, Heisenberg no logró reunir las cantidades de uranio necesarias, que erróneamente evaluó en toneladas, y pensó que se necesitarían muchos años para fabricar una bomba. Finalmente, Hitler decidió dar prioridad a los proyectos de corto plazo y Alemania renunció a la bomba atómica. Después de la guerra surgirá el interrogante de si Heisenberg planificó el fracaso del plan atómico alemán, como sostuvo él mismo, o sencillamente fracasó. Lo cierto es que, como dice Closet: "La bomba nazi nunca fue más que un fantasma de exiliados [...] Einstein había sido un juguete de la historia". En una carta a su amigo Linus Pauling, Einstein escribió que el día que firmó la carta al presidente Roosevelt había cometido el gran error de su vida. Refiriéndose a la bomba atómica nazi, escribió: "Si hubiera sabido que ese temor no se justificaba, ni yo ni Szilard hubiéramos colaborado en la apertura de esa caja de Pandora. Pues mi desconfianza de los gobiernos no se limitaba a Alemania" (Closets, 2005: 443, 449).

**Un físico pacifista en la era atómica**

Al día siguiente de la segunda explosión atómica, en la ciudad japonesa de Nagasaki, el 12 de agosto de 1945, Einstein aclaró públicamente que él nunca había trabajado en la bomba atómica: "Yo no he trabajado en este tema, en absoluto. Yo estoy interesado en la bomba igual que cualquier otra persona; probablemente un poco más interesado. Sin embargo, no me siento justificado a decir nada acerca de esto" (Lewis, 1945). No importa cuánto enfatizara Einstein este punto, el general Leslie Groves, líder militar del proyecto Manhattan, y sus consejeros, antes de la primera explosión atómica en Hiroshima ya habían redactado un breve documento donde por primera vez se relataba una versión del desarrollo del proyecto Manhattan. Este folleto, dado a conocer el 13 de agosto, ubicaba el origen científico y político de la bomba atómica en dos hechos que tenían a Einstein como protagonista: la ecuación $E = mc^2$ y la carta de Einstein al presidente Roosevelt. A partir de este momento ya no tendrán importancia los intentos de poner en orden los hechos y de atribuir responsabilidades, el mito que vincula a Einstein con la bomba atómica ya estaba en marcha. Entre los muchos ejemplos, mencionemos solo dos. El primero es un artículo publicado en el *New York Times* a fines de septiembre titulado "Bomba atómica basada en la teoría de Einstein", donde se afirmaba que: "La existencia de la energía atómica fue descubierta por Einstein alrededor de cuarenta años atrás" (Laurence, 1945). El segundo ejemplo es la portada del 1° de julio de 1946 de la revista *Time*, donde aparece Einstein y su ecuación sobre un fondo de hongo atómico. El resultado fue que la historia oficial se cristalizó en relatos del tipo que fuera publicado en *Scientific Monthly* con el título "La teoría de la relatividad y la bomba atómica", el cual comienza sosteniendo que el interés en la teoría de Einstein "se ha intensificado en los últimos meses por su conexión con la bomba atómica" (Gould, 1947: 48).

Por esos días, mientras Teller se entregaba por completo a la carrera por la bomba de fusión y la marina de los Estados Unidos realizaba dos pruebas atómicas en el atolón de Bikini, algunos físicos mostraban activamente su oposición al proceso de militarización de la física que se prolongaba de manera inexorable a los primeros años de la guerra fría. Szilard se inclinaba hacia la biología y el estudio de los delfines, Fermi a los rayos cósmicos, Wigner a la física cuántica. A comienzos de agosto de 1946 fue creado el Emergency Committee of Atomic Scientists. Presidido por Einstein e integrado por otros siete científicos que estuvieron involucrados en el desarrollo de la física atómica. Su propósito era alertar al mundo del peligro nuclear: "negarse a colaborar en asuntos militares debería ser un principio moral esencial de todos los sabios verdaderos". Este comité también sostuvo: "No hay solución al problema, excepto el control internacional de la energía atómica y, en última instancia, la eliminación de la guerra" (*Wall Street Journal*, 1946).[1] En este escenario convulsionado, mientras que Einstein aparecía públicamente como quien habría dado los primeros pasos hacia la bomba, en círculos más especializados dentro de los sectores militar y político era criticado duramente por ser el principal representante de quienes argumentaban que no debería haberse utilizado la bomba contra una población civil. Un ejemplo de estas críticas puede verse en la revista *Military Affairs*, donde se defiende la decisión de usar la bomba en Hiroshima y se acusa a Einstein de esgrimir argumentos cuyo poder de persuasión "descansa no solamente en su afinidad con nuestra visión ética básica sino también en su carácter hipotético", aunque "una demostración de la validez de la alternativa sugerida es, sin embargo, imposible" (Winnacker, 1947: 26).

Los últimos años de la vida de Einstein transcurrieron en un contexto dominado por la paranoia anticomunista de la sociedad norteamericana, motivada en parte por el paso de China a un régimen socialista y por la primera prueba nuclear de la Unión Soviética en Semipalatinsk, en agosto de 1949. Como respuesta, en los Estados Unidos se inició la construcción de un arsenal nuclear y se aceleró la sucesión de pruebas atómicas (en Nevada se realizarían más de cien ensayos). El 31 de enero de 1950 el presidente Harry S. Truman anunció su decisión de acelerar el desarrollo de un arma termonuclear. A los pocos días, en un diálogo por radio con Eleanor Roosevelt, Einstein sostuvo: "La aniquilación universal se perfila con nitidez creciente como término del proceso [...] Si no es realista la idea de un gobierno mundial, entonces la única visión realista de nuestro porvenir es la aniquilación total del hombre por el hombre". También durante esos días, mientras Einstein se manifestaba en contra de la guerra de Corea y sostenía que la carrera armamentista entre Estados Unidos y la Unión Soviética "originalmente concebida como una medida preventiva, asume un carácter histérico" (Einstein, 1950), el senador Joseph McCarthy comenzaba la brutal caza de comunistas dentro de los Estados Unidos.

---

[1] Los otros integrantes del comité fueron: Harold Urey (Universidad de Chicago), Hans Bethe (Universidad de Cornell), Thorfin Hogness (Universidad de Chicago), Philip Morse (director del Laboratorio Nacional de Brookhaven), Leo Szilard (Universidad de Chicago), Víctor Weisskopf (Instituto de Tecnología de Massachusetts), Linus Pauling (Instituto de Tecnología de California).

En un escenario superpoblado de juicios por espionaje, las intervenciones públicas de Einstein desencadenaron la intervención del FBI. El director del Federal Bureau of Investigation y líder anticomunista Edgar Hoover ordenó personalmente la investigación sobre la base de una única acusación no corroborada proveniente de una fuente no revelada que sostenía que a fines de los años veinte y comienzos de los treinta la oficina de Einstein en Berlín había sido utilizada como mensajería de agentes soviéticos. Nada de esto fue nunca probado, aunque el expediente de Einstein llegó a superar las mil quinientas páginas (Schwartz, 1989: 281). En enero de 1953, Einstein escribió (infructuosamente) a Truman para que conmutara la sentencia de muerte a Julius y Ethel Rosenberg, acusados de transmitir información sobre la bomba atómica a los soviéticos. Durante este mismo año, en una carta dirigida a un profesor secundario que se había negado a testimoniar ante el subcomité de seguridad interna del Senado (luego publicada por la prensa), Einstein expresaba:

> Cada intelectual que es llamado ante uno de estos comités debe rehusarse a testificar, esto es, debe estar preparado para la cárcel y la ruina económica, en pocas palabras, para el sacrificio del bienestar de su persona en beneficio de los intereses del bienestar de la cultura de su país (citado en Pais, 1994: 238).

Esta carta provocó un torrente de indignación pública, que incluyó acusaciones de irresponsabilidad y expresiones como "enemigo de América". En junio de 1954 Einstein se comprometió con el "caso Oppenheimer" y a comienzos de 1955 firmó el manifiesto promovido por el filósofo y matemático británico Bertrand Russell, conocido como manifiesto Russell-Einstein, que se publicará recién el 9 de julio y que pondrá en marcha el "movimiento Pugwash", todavía vigente, a partir de la conferencia que se celebrará en la ciudad de Canadá que lleva ese nombre en 1957, dos años más tarde de la muerte de Einstein ocurrida el 18 de abril de 1955.

**Referencias**